\documentclass[sigconf]{acmart}
\renewcommand\footnotetextcopyrightpermission[1]{} 
\usepackage{geometry}
\usepackage{array}
\usepackage{multirow}
\usepackage{longtable}
\usepackage{cleveref}
\usepackage{colortbl}
\usepackage{booktabs}
\usepackage{tabularx}

\AtBeginDocument{%
  }

\setcopyright{acmlicensed}
\copyrightyear{2018}
\acmYear{2018}
\acmDOI{XXXXXXX.XXXXXXX}

\acmConference[PRE-PRINT]{PRE-PRINT}{Carnegie Mellon University}{2025}

\acmISBN{978-1-4503-XXXX-X/18/06}

\begin{document}

\title{Level Up Peer Review in Education}
\subtitle{Investigating genAI-driven Gamification system and its influence on Peer Feedback Effectiveness}

\author{Rafal Wlodarski}
\email{rafal.wlodarski@sv.cmu.edu}
\affiliation{%
  \institution{Carnegie Mellon University Silicon Valley}
  \streetaddress{NASA Research Park, 23 Moffett Field}
  \city{Mountain View}
  \country{USA}
}

\author{Leonardo da Silva Sousa}
\email{leo.sousa@sv.cmu.edu}
\affiliation{%
  \institution{Carnegie Mellon University Silicon Valley}
  \streetaddress{NASA Research Park, 23 Moffett Field}
  \city{Mountain View}
  \country{USA}
}

\author{Allison Connell Pensky}
\email{aconnel2@andrew.cmu.edu}
\affiliation{%
  \institution{Carnegie Mellon University}
  \streetaddress{4765 Forbes Ave, Tepper Quad}
  \city{Pittsburgh}
  \country{USA}
}
\renewcommand{\shortauthors}{Wlodarski et al.}

\begin{abstract}
In software engineering (SE), the ability to review code and critique designs is essential for professional practice. However, these skills are rarely emphasized in formal education, and peer feedback quality and engagement can vary significantly among students. This paper introduces Socratique, a gamified peer-assessment platform integrated with Generative AI (GenAI) assistance, designed to develop students’ peer-review skills in a functional programming course. By incorporating game elements, Socratique aims to motivate students to provide more feedback, while the GenAI assistant offers real-time support in crafting high quality, constructive comments. To evaluate the impact of this approach, we conducted a randomized controlled experiment with master’s students comparing a treatment group with a gamified, GenAI-driven setup against a control group with minimal gamification. Results show that students in the treatment group provided significantly more voluntary feedback, with higher scores on clarity, relevance, and specificity - all key aspects of effective code and design reviews. This study provides evidence for the effectiveness of combining gamification and AI to improve peer review processes, with implications for fostering review-related competencies in software engineering curricula.
\end{abstract}

\maketitle

\section{Introduction}
In modern software engineering practice, peer review is widely recognized as a critical mechanism to ensure software quality and improve team collaboration \cite{bacchelli2013modern, badampudi2023modern}. Reviews allow engineers to critique each other’s code or designs, catching errors, improving maintainability, and spreading knowledge throughout the team \cite{rigby2017convergent, indriasari2020gamification}. However, recent studies have highlighted significant gaps between educational outcomes and industry expectations, with feedback emerging as one of the most critical non-technical skills \cite{groeneveld2021acm}. In fact, the students themselves express a strong desire to improve their feedback-giving abilities \cite{hazzan2014teaching}, recognizing their value both in academic settings and in the workplace.
Traditionally, teachers have been seen as the primary source of feedback in the classroom. However, the increasing complexity and size of classrooms have highlighted the challenges and impracticality of teachers providing personalized feedback to all students on every task. Thus, peer feedback, where students assess or evaluate each other's work, can help manage the administrative load in large classes and can inform summative assessments \cite{luo2014peer}. More importantly, by practicing giving and receiving constructive criticism, students can develop the feedback literacy skills necessary for their future careers~\cite{wang2012assessment, politz2016sweep, kubincova2017code}.

Motivated by these insights, we developed \textit{Socratique}, a platform, to enhance the peer feedback process. Our approach aims to boost student engagement and teach them how to craft feedback that parallels professional SE review standards. Central to this platform is the use of gamification and Generative Artificial Intelligence. Gamification, defined as the integration of game-like elements into non-game contexts \cite{deterding2011gamefulness}, aims to boost intrinsic motivation (e.g. becoming a better reviewer) through extrinsic rewards (e.g., earning points, badges). Additionally, \textit{Socratique} builds on  Generative Artificial Intelligence (GenAI), a subset of AI systems that is designed to create new content or generate data that mimic human output. We integrated a GenAI assistant ("Mr. Pepper") into the platform, to assist students in real-time on how to provide constructive and meaningful insights to their peers.
We hypothesize that an approach that integrates gamification and a GenAI assistant will motivate students to submit more high-quality feedback. 

The contribution of this paper is threefold. First, we present \textit{Socratique}, a gamified peer feedback platform that integrates Generative AI (GenAI) to help students refine their feedback. We describe the design of Socratique, highlighting how its gamified mechanics were crafted to boost engagement in peer review. Second, we investigate the impact of the platform on peer feedback effectiveness through a controlled experiment conducted with master’s students in a functional programming course. Finally, we discuss the implications for software engineering education: examining how gamification can enhance the peer review process and how GenAI can be a scaffolding tool for higher-quality, formative feedback.

In the next sections, we present our GenAI-driven gamified approach (\Cref{socratique}) and the controlled experiment (\Cref{experiment_design}). Our findings, outlined in sections 5 and 6,  suggest that both the quantity and quality of peer feedback increase when students engage with gamification that includes GenAI assistance.

\section{Related work}
Peer feedback systems are a practical means to simulate real code or design review processes in classroom settings \cite{indriasari2020gamification}. These can either be off-the-shelf products \cite{turner2018peer, kubincova2017code} like Bitbucket, Github, Phabricator, Codebrag, Upsource or software solutions tailor-made  for that purpose \cite{li2012facilitating, hundhausen2013talking, wang2015toward}. The bespoke tools implement features meant to support the peer code review process - namely anonymous reviews, rubrics-based assessment, visualization of code history, or distribution of reviews. Such tools suffered from certain barriers that prevented their effective usage. In fact, many studies reported low learning engagement that manifests itself through low enthusiasm, a low level of participation, and rushed reviews \cite{hundhausen2010design, turner2011student, hundhausen2011online, sripada2015code}. Another key challenge to the adoption of peer feedback practices in educational contexts is the low quality and trustworthiness of reviews provided by students \cite{turner2008misunderstandings, ohara2016incorporating}.

An approach often used to foster student participation in peer review is gamification. By introducing leaderboards, badges, or rewards, educators aim to increase students’ motivation to provide and engage with peer feedback. However, a recent review of the literature on gamified peer review in education \cite{indriasari2020gamification} indicates that while 13 studies report a positive impact on student enthusiasm and participation, 5 studies found neutral effects. These mixed results highlight an incomplete picture of the true efficacy of gamification and suggest the need for further investigation of when, how, and why gamification strategies succeed or fail in improving peer feedback effectiveness. This is in line with another recent review of the literature on peer code evaluation in higher education \cite{indriasari2020review}.

The second key challenge to the adoption of gamification in peer feedback, the low quality of reviews, was historically difficult to address in an automated way \cite{indriasari2020gamification}. In fact, they call for qualitative and subjective metrics, such as the opinions of peers or instructors, to assess the quality of the provided critiques \cite{indriasari2020gamification}. However, with Generative AI which is increasingly recognized as a valuable tool for providing automated feedback in educational contexts, it is no longer an obstacle. Their capacity to process textual inputs at scale makes them especially well suited for tasks such as qualitative evaluations of textual critiques. For example, Bernius et al. \cite{bernius2022machine} demonstrate how LLM-driven summative feedback can lighten instructors’ grading workload for text-based problem-solving exercises. Nguyen et al. \cite{nguyen2023evaluating} extend this approach to more open-ended, gamified assignments, where generative models offer insight into student performance. Meanwhile, Han et al. \cite{han2023fabric} propose a system that provides rubric-oriented scoring and formative suggestions for longer textual submissions (e.g., student essays). Although many of these studies focus on textual input, they underscore the potential of LLM-based feedback to enhance educational processes, particularly in domains such as software engineering, where students’ work often requires a detailed critique of both content and structure. 

Although gamification has been explored in feedback sharing systems before and Generative AI recently used in the feedback provision process, our platform is distinct due to its integration of a real-time feedback assistant powered by GenAI, which can enhance the effectiveness of gamification elements. To our knowledge, this study is the first to combine generative AI with gamification in feedback systems, particularly for software engineering education.

\section{Socratique: A Gamified Platform}
\label{socratique}

In an effort to build feedback literacy skills among graduates of the Software Engineering program at our university, we implemented a bespoke web-based peer-feedback platform called \textit{Socratique} (a portmanteau of the words `\textit{Socrates}' and `\textit{critique}'). We designed it to improve the peer feedback process through the integration of \textit{gamification} elements and a \textit{Generative  Artificial  Intelligence} (GenAI) tutor. We built the platform to foster student participation in feedback activities, making the feedback process more interactive and educationally enriching. The platform was piloted in Spring 2024 in the context of one of the program's core courses - Functional Programming in Practice. Though piloted in a functional programming context, the platform is readily applicable to any SE course where students submit code, project documentation, or design diagrams for peer evaluation. Peer reviewers can therefore emulate real peer review tasks by commenting on syntax, readability, algorithmic complexity, or design trade-offs.

\subsection{Platform Overview}
By combining the gamification elements with GenAI, \textit{Socratique} aims to make peer feedback not only more accessible but also more valuable. It encourages students to engage with their peers' work in a thoughtful way and cultivate critical feedback skills.

\textbf{Features for Instructors}: From an instructor's perspective, \textit{Socratique} serves as a comprehensive tool for managing courses, assignments, and peer feedback activities. It provides functionality to create and manage assignments, design questionnaires, and distribute student submissions to students for review. Instructors can either create questionnaires from scratch or use saved templates, offering a degree of question types, such as open-ended, multiple-choice, Likert scale, and rating scales, allowing flexibility in structuring feedback and assessments. The platform also supports the distribution of deliverables (any digital output such as presentations, documentation, or source code) among students, ensuring that peer review is handled efficiently.

\textbf{Features for Students}: For students, \textit{Socratique} acts as a centralized venue to submit their deliverables and participate in anonymous peer feedback. The platform allows students to evaluate their peers' work, view feedback on their submissions, and interact with summative statistics from structured questions (such as ratings or Likert scale questions). Furthermore, \textit{Socratique} offers a clarification feature that allows students to request further explanation or details on the feedback they have received, facilitating deeper understanding and dialogue between the reviewers.

\textbf{Gamification}: To encourage active participation, \textit{Socratique} incorporates gamification mechanics (\Cref{tab:MDAframework}), which are crucial to engaging students in the feedback process. Students can earn points, badges, and rewards based on their interactions with the platform, such as completing reviews or providing feedback (Table~\ref{tab:feedback_rubric}). Gamification elements such as leader boards and badges motivate students to provide timely feedback and improve its quality. In peer review contexts, thoroughness and consistency can wane over time. Our gamification elements (points, badges, leaderboards) incentivize students to sustain engagement across multiple review cycles, reflecting the iterative nature of software development.

\textbf{GenAI assistance}: One of \textit{Socratique}'s key innovations is its integration of Generative Artificial Intelligence (GenAI), through a virtual tutor named Mr. Pepper. The tutor offers real-time guidance on improving the open-ended feedback that students provide to their peers. The GenAI-driven tutor helps students craft more constructive, meaningful, and actionable feedback, fostering feedback literacy. Our assistant emphasizes communication clarity, detail, and relevance by scaffolding the human aspect of peer review. By consulting Mr. Pepper, students can receive suggestions for refining their critiques, aligning them with best-practices of quality feedback, making the feedback sharing process more formative, and ultimately preparing students to succeed in their career.

\textbf{Social Interaction Features}; \textit{Socratique} also includes a social interaction feature called poking, inspired by Facebook's early mechanism for nudging friends \cite{facebookpoke}. Through this feature, students can nudge their peers to provide a pending review, gently reminding them to meet their feedback obligations. This feature helps maintain a level of peer accountability and keeps the feedback cycle moving smoothly, particularly in large class settings.

\begin{figure*}[h]
    \centering
    \includegraphics[width=\textwidth]{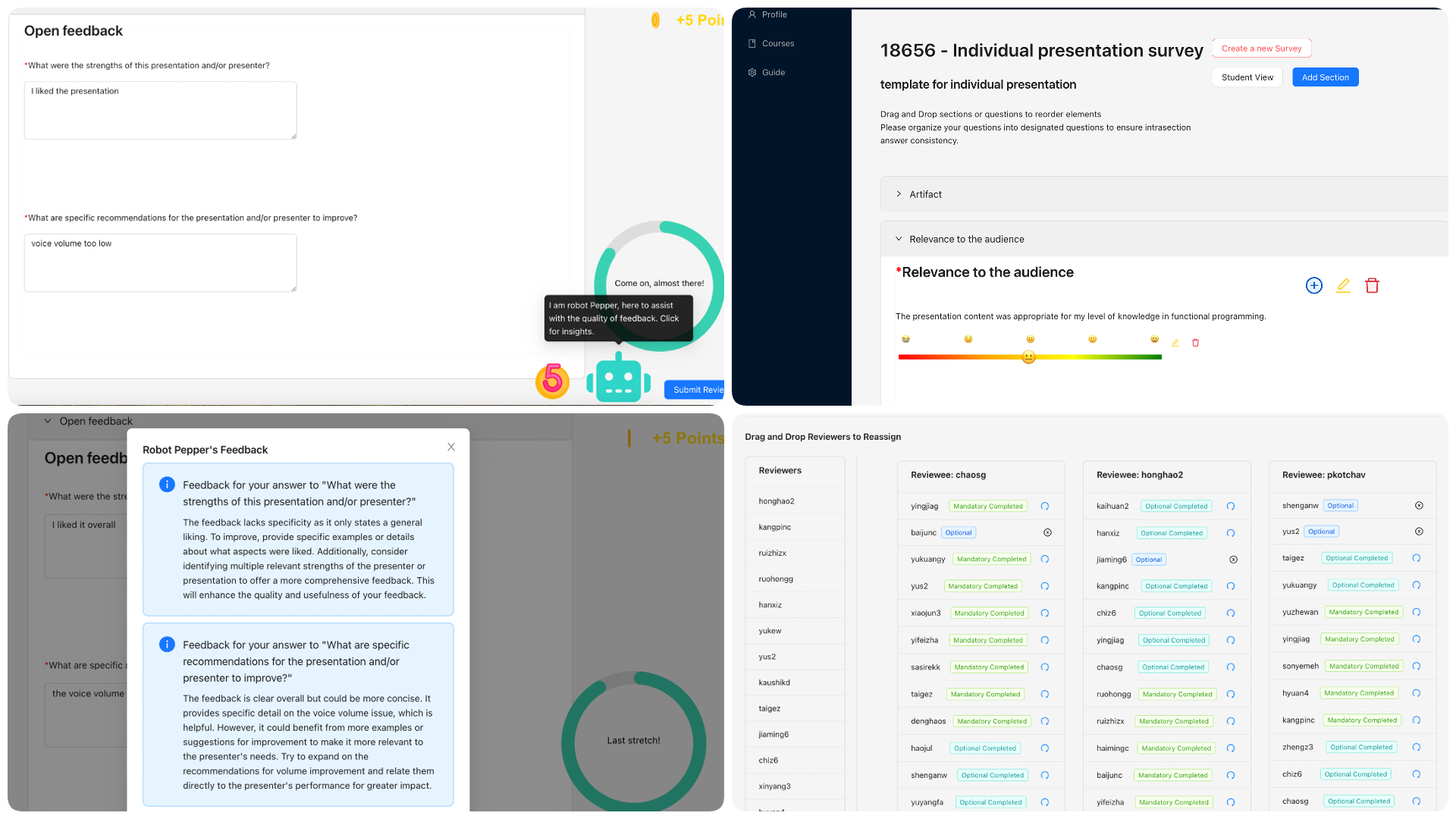}
    \caption{An overview of the Socratique platform from top-left clockwise: 1) GenAI assistant "Mr. Pepper" 2) set up of a survey template 3) peer reviews assignment dashboard 4) GenAI assistant's feedback }
    \label{fig:socratiquePlatfrom}
\end{figure*}

\subsection{Use of Gamification}
\label{subsec:sub22}

To introduce a meaningful and coherent gamification approach, we adopted the well-recognized mechanics-dynamics-aesthetics (MDA) framework \cite{hunicke2004mda}. It allowed us to align the emotional responses that we wanted to evoke in the students with specific game mechanics designed to achieve our objectives (\Cref{tab:MDAframework}). 

\begin{table}[]
\resizebox{\columnwidth}{!}{%
\begin{tabular}{l|l|l}
\hline
\rowcolor[HTML]{9B9B9B} 
\multicolumn{1}{c|}{\cellcolor[HTML]{9B9B9B}{\color[HTML]{333333} \textbf{\begin{tabular}[c]{@{}c@{}}Gamification\\ objective\end{tabular}}}} & \multicolumn{1}{c|}{\cellcolor[HTML]{9B9B9B}{\color[HTML]{333333} \textbf{Dynamics}}} & \multicolumn{1}{c}{\cellcolor[HTML]{9B9B9B}{\color[HTML]{333333} \textbf{Mechanics}}}                                                                                                                                                                                                        \\ \hline
                                                                                                                                              & Rivarly                                                                               & \begin{tabular}[c]{@{}l@{}}Points awarded for desired actions :\\   - provide a mandatory review in a timely manner\\   - provide an optional review in a timely manner \\   - provide a mandatory review past the deadline\\   - provide an optional review past the deadline\end{tabular} \\ \cline{2-3} 
                                                                                                                                              & Rivalry                                                                               & Leaderboard (display the rankings of participants)                                                                                                                                                                                                                                           \\ \cline{2-3} 
                                                                                                                                              & Reward                                                                                & \begin{tabular}[c]{@{}l@{}}Store (exchange of experience points for real-world\\ prizes like bonus course points, final exam question\\ waiver)\end{tabular}                                                                                                                                 \\ \cline{2-3} 
                                                                                                                                              & Time pressure                                                                         & Countdown of mandatory and optional surveys left                                                                                                                                                                                                                                             \\ \cline{2-3} 
\multirow{-5}{*}{\begin{tabular}[c]{@{}l@{}}Encourage the users\\ to come back\\ and provide \\ more feedback\\ than required\end{tabular}}       & Lottery                                                                               & \begin{tabular}[c]{@{}l@{}}Roulette (draw a roulette to obtain additional points\\ in the next survey)\end{tabular}                                                                                                                                                                          \\ \hline
                                                                                                                                              & Achievement                                                                           & \begin{tabular}[c]{@{}l@{}}Badges (granted to users based on achievement of\\ specific quality targets)\end{tabular}                                                                                                                                                                         \\ \cline{2-3} 
                                                                                                                                              & Reward                                                                                & Point multipliers awarded along with specific badges                                                                                                                                                                                                                                         \\ \cline{2-3} 
\multirow{-3}{*}{\begin{tabular}[c]{@{}l@{}}Encourage the\\ users provide\\ feedback of \\ higher quality\end{tabular}}                       & Rivalrly                                                                               & \begin{tabular}[c]{@{}l@{}}Points awarded for desired actions:\\   - consulting GenAI tutor for the first time\\   - consulting GenAI tutor when open-ended feedback\\ score is low (\textless{}6)\end{tabular}                                                                        \\ \hline
\end{tabular}%
}
\caption{ Gamification system (with MDA framework)}
\label{tab:MDAframework}
\end{table}

Our primary goal was to foster a sense of competition among students, encouraging them to provide numerous high-quality peer reviews. To accomplish this, we incorporate several gamification elements \cite{alhammad2018gamification}, including points, progress indicators, leaderboards, and rewards. Students could earn experience points (XP) by performing desirable actions, such as submitting peer reviews or providing high-quality feedback. Students could use these XP to redeem rewards on the platform. The instructor maintained complete control over the definition of available rewards, their cost of XP and their quantity, allowing customization that improved student motivation.

In addition to competition, our aim was to instill a sense of achievement. To do this, we introduced a series of badges that students could earn based on the quality of their reviews. Quality was evaluated using our rubric (\Cref{tab:feedback_rubric}) on a 9-point scale, and we established three quality thresholds for badges: scores greater than 6, 7, and 8. Each threshold was associated with a unique badge: "Curious Commentor," "Comment Captain," and "Comment Crusader," respectively. Furthermore, within each badge type, three levels were introduced: bronze, silver, and gold, corresponding to 1, 3, and 6 peer reviews that met the quality threshold. The silver and gold badges came with a multiplier effect, increasing the XP earned for subsequent reviews and providing additional incentives for students to continuously improve the quality of their feedback. The badges were not mutually exclusive and the students were notified whenever they earned a new badge. We added badge icons to an accomplishments bar that was always visible to students.

To further boost engagement and motivate students to provide additional, non-mandatory reviews, we introduced an element of surprise through a lottery system. This was visualized as a Prize Wheel that became accessible once the students had completed their mandatory reviews. The Prize Wheel featured sections containing preconfigured rewards with varying levels of desirability - from 0 to 15 XP — and different probabilities of winning each prize. Students could spin the wheel to determine the bonus XP they would receive for completing their next optional review. The lottery could be entered for every optional review, adding a layer of gamified excitement and encouraging further participation.

All gamification mechanics and rules were clearly documented and accessible within the \textit{Socratique} platform to ensure transparency and support intended usage. Combining these elements, we believe that we created a stimulating environment that nurtured curiosity, encouraged active participation, and promoted continuous improvement through healthy competition and intrinsic motivation.

\subsection{Use of Generative AI}
\label{subsec:sub23}

To enhance the peer feedback process within \textit{Socratique}, we integrated Generative AI (GenAI), specifically leveraging OpenAI's ChatGPT-3.5-turbo model to support key gamified actions (e.g., answering open-ended questions). This AI component, embodied as a virtual assistant named Mr. Pepper, serves two primary functions:
\begin{itemize}
    \item Assist students in improving the quality of their open-ended feedback.
    \item Evaluate the quality of open-ended feedback in real time.
\end{itemize}

An icon of a robot representing the GenAI tutor was placed alongside open questions. Upon request, Mr. Pepper would analyze the content of open-ended feedback fields and evaluate it based on a preconfigured rubric (Table~\ref{tab:feedback_rubric}) which outlines criteria for effective peer reviews. The AI then provided feedback to students, highlighting the strengths of their critique and offering suggestions for improvement. Students could incorporate the AI's suggestions and consult Mr. Pepper as often as needed. However, to promote exploration of the tool without overreliance, points were awarded only for the first use of the tutor during each review session.

Upon submission of the peer evaluation questionnaire, the tutor is automatically activated to assess the overall quality of the feedback using the same rubric\footnote{We do not apply the assistance to the content of the peer-reviewed artifact (e.g., a PDF presentation)}. If the feedback met specific quality thresholds, the student earned a badge and received a point multiplier.
If a student's feedback quality score fell below 50\%, a pop-up window prompted them to request assistance from the GenAI tutor. In this scenario, students received additional points for requesting GenAI's help, though this reward was limited to one time per poor-quality review within a review session.

Mr. Pepper was not re-trained or fine-tuned for peer feedback tasks in this study. Instead, we employed strategic prompt engineering to align GenAI's responses with our educational objectives. These tailored prompts guide AI to focus on the substantive aspects of feedback, such as clarity, relevance, and specificity, ensuring that the assistance provided is contextually appropriate and pedagogically meaningful. This approach mitigates the risk of GenAI simply improving text legibility by directing it to foster constructive criticism and actionable insights, thus enhancing the quality and effectiveness of peer reviews in software engineering education.

To ensure that our prompts guided GenAI to provide relevant and constructive feedback, we conducted preliminary validations using peer feedback submissions from previous course iterations. These submissions were reviewed and annotated by the course instructor and co-authors, allowing us to refine the prompts to align with the educational objectives and the specific criteria in our rubric.

\begin{table*}[]
\centering
\begin{tabular}{>{\raggedright\arraybackslash}p{6cm}|>{\raggedright\arraybackslash}p{2.5cm}|>{\raggedright\arraybackslash}p{8cm}}
\hline
\rowcolor[HTML]{9B9B9B} 
\textbf{Criterion} & \textbf{Score} & \textbf{Description} \\ 
\hline

\multirow{4}{*}{\begin{tabular}[c]{@{}l@{}}Feedback Clarity Score: The feedback should be\\ easy to understand, express ideas clearly, and\\ avoid ambiguity or verbosity.\end{tabular}} 
& \textbf{Exemplary (3)} & The feedback is exceptionally clear, with precise language and well-structured ideas that are easy to understand. \\ \cline{2-3} 

& \textbf{Proficient (2)} & The feedback is clear overall, but may contain minor ambiguities or could be slightly more concise. \\ \cline{2-3} 

& \textbf{Developing (1)} & The feedback communicates its points but could benefit from additional simplification or organization to enhance clarity. \\ \cline{2-3} 

& \textbf{Unsatisfactory (0)} & The feedback lacks clarity, containing significant ambiguity or confusion that hinders understanding. \\ 
\hline

\multirow{4}{*}{\begin{tabular}[c]{@{}l@{}}Feedback Relevance Score: The feedback should\\ bear significance to the context (e.g., individual\\ presentation, content knowledge, audience\\ engagement, delivery, clarity).\end{tabular}} 
& \textbf{Exemplary (3)} & 4-5 relevant strengths and/or weaknesses were identified. \\ \cline{2-3} 

& \textbf{Proficient (2)} & 2-3 relevant strengths and/or weaknesses were identified. \\ \cline{2-3} 

& \textbf{Developing (1)} & 1 relevant strength or weakness was identified. \\ \cline{2-3} 

& \textbf{Unsatisfactory (0)} & No relevant strengths or weaknesses were identified. \\ 
\hline

\multirow{4}{*}{\begin{tabular}[c]{@{}l@{}}Feedback Specificity Score: The feedback \\ should help the recipient understand \\ exactly what aspects of their work \\ are being addressed.\end{tabular}} 
& \textbf{Exemplary (3)} & Specific examples or details were given for 3-5 strengths and/or weaknesses identified. \\ \cline{2-3} 

& \textbf{Proficient (2)} & At least 2 strengths and/or weaknesses had specific details or examples. \\ \cline{2-3} 

& \textbf{Developing (1)} & At least 1 strength or weakness had specific details or examples. \\ \cline{2-3} 

& \textbf{Unsatisfactory (0)} & No specific example or detail was provided for strengths or weaknesses. \\ 
\hline
\end{tabular}

\caption{Feedback quality rubric}
\label{tab:feedback_rubric}
\end{table*}

\section{Study Design}
\label{study_design}

This study explores the effectiveness of a GenAI-driven gamification approach in improving the peer review process. 

\subsection{Research Questions}
\label{research_questions}

In our study, we conducted a randomized controlled experiment in the context of a master's course in functional programming to answer the following research questions.

\begin{itemize}
    \item RQ1: Does gamification increase the amount of peer feedback students provide?
    \item RQ2: Does GenAI assistance improve the quality of peer feedback?
\end{itemize}

\subsubsection{Null and Alternative Hypotheses}
\label{hypotheses}

Building on previous research on gamification in education \cite{tenorio2016gamified,demarcos2017social,khandelwal2017impact,hsu2018game,rojas2017code} and the rise of Generative AI, we investigate the following hypotheses.

\begin{itemize}
    \item \(H_0\): There are no significant differences in the quantity and quality of peer feedback between students using the \textit{Socratique} platform (treatment group) and those of the control group.
    \item \(H_1\): Students using the gamified peer assessment tool will voluntarily provide \textit{more feedback}, as measured by the average number of peer reviews submitted, compared to the control group.
    \item \(H_2\): Students using the gamified peer assessment tool will provide \textit{ higher quality feedback}, measured in terms of clarity, relevance, and specificity in open responses, compared to the control group.
\end{itemize}

\subsubsection{Statistical Test}

This study aims to compare the quantity and quality of peer feedback between 
students using the \textit{Socratique} platform (treatment group) and those 
in the control group (using \textit{Socratique} without gamification elements). 
To achieve this, we used a mixed-model Analysis of Variance (ANOVA)—with 
Session (1 vs. 2) as a within-subject factor and Condition (treatment vs. control) 
as a between-subjects factor—and independent sample t tests to determine 
whether the observed differences between groups were statistically significant, 
using an alpha level of 0.05.

\textbf{Hypothesis Testing:}
\begin{itemize}
    \item \textbf{RQ1 - Quantity of Feedback (\(H_1\))}: We conducted a mixed-model ANOVA to evaluate the main effects of time (Session 1 vs. Session 2) and condition (control vs. treatment) as well as their interaction, followed by independent-sample t tests for specific pairwise comparisons.

    \item \textbf{RQ2 - Quality of Feedback (\(H_2\))}: Similarly, we used a mixed-model ANOVA with the same within- and between-subjects factors to assess feedback quality based on clarity, relevance, and specificity, followed by t tests where needed.
\end{itemize}

These statistical tests allow us to assess whether the gamification 
and GenAI assistance provided by \textit{Socratique} lead to meaningful 
improvements in both the quantity and quality of peer feedback, 
beyond what could be attributed to random variation.

\subsubsection{Variables}
\label{variables}

In order to study the effectiveness of a GenAI-driven gamified platform, we investigated the following variables in a randomized controlled experiment.

\textbf{Independent variable}: The gamification approach implemented on the \textit{Socratique} platform, which involved incorporating game elements that could be activated or deactivated. The students were randomly assigned to the treatment condition (gamified \textit{Socratique}) or the control condition (non-gamified \textit{Socratique}).

\textbf{Dependent variables}: The peer feedback effectiveness measures included the following. 
\begin{itemize}
    \item \textbf{Feedback Quantity}: Measured by the total number of voluntary reviews submitted by each participant.
    \item \textbf{Feedback Quality}: Evaluated using a rubric focused on three facets: clarity, specificity, and relevance. These aspects are essential indicators of effective feedback, as suggested in previous research \cite{haughney2020, patchan2016} and mimic quality peer feedback in the practice of industrial software engineering.
\end{itemize}

Huang and Hew \cite{huang2018gamification} reported that the quality of the artifacts produced in an activity can be a measure of the performance of a gamification approach. Regarding peer feedback, we focused on evaluating open-ended responses and used a rubric (\Cref{tab:feedback_rubric}) created for that purpose by the instructor of the course and a member of the University Center for Teaching and Learning.

The rubric aimed at 3 facets of feedback: clarity, specificity, and relevance. The first two aspects are widely recognized as important indicators of quality feedback. Haughney et al. \cite{haughney2020} deem them necessary for a student to be able to understand how to proceed, and Pathachn et al. \cite{patchan2016} consider them as a significant predictor of student improvement. We decided to additionally consider relevance to avoid having students comment on aspects outside of the presentation's scope and to facilitate a fair and automated in-game rubric evaluation by GenAI. Rubric scores were determined by blinding each feedback entry from the treatment and control groups, randomizing the feedback, and then the instructor applied the grading rubric to each feedback submitted.

\subsection{Experiment Design}
\label{experiment_design}

We conducted a randomized controlled experiment to evaluate the influence of gamification on peer feedback. We randomly assigned the students to the control group (n = 17) or the treatment group (n = 17). Both groups used the \textit{Socratique} platform for peer review tasks. However, the treatment group used the gamified version of the platform, while the control group used a standard version. In this version, they had access to the same features (available for the treatment group, including the GenAI assistant), but not to the game elements (points, badges, etc.).

\subsubsection{Study Environment}
\label{study_environment}

The experiment took place in a 12-unit master's course on functional programming during Spring 2024, with 34 students voluntarily participating. The course included contact classes twice a week, each one hour and 50 minutes. The course content encompassed both fundamental principles of functional programming, such as first-class and higher-order functions, algebraic data types, pattern matching and recursion, and advanced concepts essential for developing realistic software systems using the functional paradigm. These advanced topics included pipeline-based big data processing, concurrent and asynchronous programming, Domain-Driven Design, and onion architecture. As part of the course, each student gave a 5-7 minute presentation on one of three predefined topics:
\begin{itemize}
    \item Functional constructs and concepts in other languages;
    \item The application of functional programming in the industry.; 
    \item Comparative analysis of functional programming versus other programming paradigms.
\end{itemize}

The presentations were structured to include an overview of the topic, practical examples, and a demonstration of functional programming principles in real-world scenarios. Presentations were distributed over two classes taking place in separate weeks, and the evaluation process followed a pre-defined timeline each time: presentations and artifact submissions on day D; peer assessment from D+1 to D+4; results visualization starting D+5. This setup was motivated by the timeliness of feedback, as research indicates that feedback is most effective when it is given promptly \cite{hattie2007power}. The presence during both classes was mandatory, but the students were allowed to choose when they would like to present. All listeners were encouraged to take notes during the presentations so that they can be shared at a later stage on the \textit{Socratique} platform. The students did not know beforehand for which presentations they would provide feedback.

Each student had to evaluate a subset of their peers after each of the presentation sessions. The platform allocated the mandatory reviews based on the number of presentations in each session (17) and the desired number of reviews each student would receive (6). This resulted in every student having three mandatory reviews to provide. They had to complete mandatory evaluations before optional reviews were available, one at a time. The number of optional reviews was capped (6) so that students could only collect a limited number of points within one evaluation period. Thus, 12 was the maximum number of reviews a student could give.

Experience points collected as part of the course (visible to the treatment group but hidden for the control group) could be exchanged at any time for the following rewards (one of each):
\begin{itemize}
    \item (300 points) Bonus course points: 4 points (\%)
    \item (250 points) Bonus course points: 2 points (\%)
    \item (200 points) Final exam question waiver (regrade of a selected question)
\end{itemize}

The number of points to collect for a given reward was derived based on the total number of points possible to obtain within the course and accounted for the random factor of the element of the "lottery" game (see Table ~\ref{tab:MDAframework}).

\subsubsection{Treatment and Control Groups}
\label{design:groups}

Students in the treatment and control groups were subject to the exact same course level requirements: each student was expected to give one presentation and complete a set number of mandatory peer reviews in a timely manner for each of the two presentation sessions. All participants followed the same process using the \textit{Socratique} platform and had access to the GenAI tutor, Mr. Pepper.

However, the control group had all gamification features (\Cref{subsec:sub22,subsec:sub23}) disabled, with these elements hidden from the user interface. Consequently, control group students did not receive explicit incentives to provide optional peer reviews (such as collecting points redeemable for rewards), nor were they explicitly encouraged to improve feedback quality through interactions with GenAI. Furthermore, they were not exposed to elements that promote competition (e.g., leaderboards) or achievement (e.g., badges and point multipliers). This distinction allows us to isolate the effects of gamification and GenAI on the quantity and quality of peer feedback.

It should be noted that, at the end of the semester, the same rewards were distributed to the participants in the control group based on the gamification points that would have been earned (tracked invisibly). These rewards were assigned to the top three scorers in the control group to ensure equitable treatment while still selectively incentivizing desired behaviors only among the students who received the gamification treatment.

\subsection{Experiment Procedure}
\label{operations}

\subsubsection{Preparation}
\label{preparation}

We established the study's design in advance and it was reviewed by all authors. Prior to the experiment, a kick-off session was conducted to inform students of the educational research details, emphasizing that choice of whether or not to voluntarily participate would not affect their course grade. The study was approved by the university’s Institutional Review Board (STUDY2016\_00000148). To ensure consistency, a training session was held to guide students in using the platform and providing constructive peer feedback.

\subsubsection{Execution}
\label{execution}

During the first presentation session, all students received a walk-through of the platform that covered aspects such as UI navigation and review processes. We also provided guidelines on giving effective feedback and the students completed a quiz on best practices.

\subsubsection{Data Collection}
\label{data_collection}

Data collection involved two metrics:
\begin{itemize}
    \item \textbf{Feedback Quantity}: The data were derived from the number of voluntary reviews recorded in the platform database.
    \item \textbf{Feedback Quality}: Evaluated by GenAI in real-time using a rubric to generate quality badges, and reassessed post-experiment by the instructor using the same rubric. This data was used to answer RQ2. In addition, a questionnaire was distributed at the end of the semester to gather feedback on platform usage.
\end{itemize}

\subsubsection{Post-Experiment Survey} Although it was not formally part of the research questions, we asked students to complete a survey at the end of the semester to provide feedback about their experience with the \textit{Socratique} platform. This survey aimed to collect additional information on their interaction with various features of the platform, including gamification elements and the GenAI assistant, \textit{Mr. Pepper}. By collecting this feedback, we sought to better understand students' perceptions of the platform's usability and effectiveness in enhancing the peer feedback process, which could help inform future iterations of the tool.

\subsubsection{Replication and Tool Accessibility}
To facilitate replication and further research, the \textit{Socratique} platform will be made available to interested researchers on request. In addition, the source code is hosted on a public repository (GitHub) to ensure transparency and ease of access for replication studies.

\section{Quantitative Results}

For the experiment, the students presented two sessions in separate weeks (\Cref{study_environment}). Therefore, to answer our research questions, we collected data at two points in time: Session 1 (S1) and Session 2 (S2). Each session corresponds to the presentation of half of the students. We evaluated the quality of open peer feedback using a rubric that assessed clarity, relevance, and specificity, with each category rated on a scale of up to 3 points (\Cref{tab:feedback_rubric}). A single coder (the instructor) evaluated the feedback and was blinded to the conditions during the scoring. The quantity of feedback provided by students was measured using a simple count, with any feedback greater than 6 within a session being optional.

The summarized results are presented in~\Cref{tab:feedback_comparison}. We have highlighted in light gray (column 4) the results that are statistically significant using the significance level 5\% ($\alpha = 0.05$). 

\begin{table}[]
\resizebox{\columnwidth}{!}{%
\begin{tabular}{c|r|r|l}
\hline
\rowcolor[HTML]{9B9B9B} 
Survey Item & \multicolumn{1}{c|}{\cellcolor[HTML]{9B9B9B}\begin{tabular}[c]{@{}c@{}}Control\\ Mean\end{tabular}} & \multicolumn{1}{c|}{\cellcolor[HTML]{9B9B9B}\begin{tabular}[c]{@{}c@{}}Treatment\\ Mean\end{tabular}} & \multicolumn{1}{c}{\cellcolor[HTML]{9B9B9B}\begin{tabular}[c]{@{}c@{}}Statistical\\ Result\end{tabular}} \\ \hline
\textbf{\begin{tabular}[c]{@{}c@{}}Amount of feedback given across both sessions\\ (min 12 - number of required peer reviews)\end{tabular}} & 16.9 & 23.0 & \cellcolor[HTML]{C0C0C0}\begin{tabular}[c]{@{}l@{}}\textit{t(23)} = -2.03\\ \textit{p} = .05\end{tabular} \\ \hline
Amount of feedback given (Session 1) & 9.6 & 13.6 & \cellcolor[HTML]{C0C0C0}\begin{tabular}[c]{@{}l@{}}\textit{t(23)} = -2.19\\ \textit{p} = .04\end{tabular} \\ \hline
Amount of feedback given (Session 2) & 7.3 & 9.5 & \begin{tabular}[c]{@{}l@{}}\textit{t(23)} = -1.28\\ p = .21\end{tabular} \\ \hline
\textbf{\begin{tabular}[c]{@{}c@{}}Overall quality of feedback given across both\\ sessions (max 9)\end{tabular}} & 4.1& 5.6& \cellcolor[HTML]{C0C0C0}\begin{tabular}[c]{@{}l@{}}\textit{t(23)} = -2.41\\  \textit{p} = .03\end{tabular} \\ \hline
Session 1: Total quality of feedback & 4.1& 5.6& \cellcolor[HTML]{C0C0C0}\begin{tabular}[c]{@{}l@{}}\textit{t(23)} = -2.41\\ \textit{p} = .03\end{tabular} \\ \hline
Session 1: Average Clarity & 2.6 & 2.6 & \begin{tabular}[c]{@{}l@{}}\textit{t(23)} = -.09\\ \textit{p} = .93\end{tabular} \\ \hline
Session 1: Average Relevance & 1.3 & 1.9 & \cellcolor[HTML]{C0C0C0}\begin{tabular}[c]{@{}l@{}}\textit{t(23)} = -4.17\\ \textit{p} < .001\end{tabular} \\ \hline
Session 1: Average Specificity & 0.3 & 1.3 & \cellcolor[HTML]{C0C0C0}\begin{tabular}[c]{@{}l@{}}\textit{t(23)} = -3.87\\ \textit{p} < .001\end{tabular} \\ \hline
Session 2: Total quality of feedback & 4.6& 5.5& \begin{tabular}[c]{@{}l@{}}\textit{t(21)}= -1.61\\ \textit{p} = .12\end{tabular} \\ \hline
Session 2: Average Clarity & 2.6 & 2.6 & \begin{tabular}[c]{@{}l@{}}\textit{t(21)}= -.24,\\ \textit{p} = .81\end{tabular} \\ \hline
Session 2: Average Relevance & 1.5 & 1.7 & \begin{tabular}[c]{@{}l@{}}\textit{t(21)}= -1.33,\\ \textit{p} = .20\end{tabular} \\ \hline
Session 2: Average Specificity & 0.5 & 1.1 & \cellcolor[HTML]{C0C0C0}\begin{tabular}[c]{@{}l@{}}\textit{t(21)}= -1.94\\ \textit{p} = .07\end{tabular} \\ \hline
\end{tabular}%
}
\caption{Quantitative Results}
\label{tab:feedback_comparison}
\end{table}

\subsection{Does incorporating gamification encourage students to provide more peer feedback?}

RQ1 examines the amount of feedback given by students using a gamified version of the \textit{Socratique} platform (treatment group) and students using a non-gamified version of the tool (control group). As shown in \Cref{tab:feedback_comparison}, students in the treatment group consistently provided more peer evaluations compared to the control group. Specifically, the students in the treatment group evaluated an average of four more peer presentations in session 1 and 2.2 more in session 2 than those in the control group.

We conducted a series of mixed model ANOVAs with time of feedback (Session 1, Session 2) as the within-subject variable and condition (control, treatment) as the between-subject factor, examining each dependent variable independently. For the measure of the quantity of feedback, the analysis revealed a significant main effect of condition (\textit{f}(1,23) = 4.11, \textit{p} = .05, $\eta_p^2$ = .15), indicating that students in the treatment group provided significantly more feedback \textit{(M} = 23.0) than those in the control group (\textit{M} = 16.9). The effect size ($\eta_p^2$ = .15) suggests a moderate impact of the intervention on the quantity of feedback, implying that the intervention successfully encouraged more active participation in peer assessment.
There was a significant main effect of time (\textit{f}(1,23) = 12.87, \textit{p} = .002, \(\eta_p^2\) = .36) such that more feedback was given during session 1 (\textit{M} = 11.3) than during session 2 (\textit{M} = 8.2). There was a significant main effect of condition (\textit{f}(1,23) = 4.11, \textit{p} = .05, \(\eta_p^2\) = .15) such that the intervention group gave more feedback (\textit{M} = 23.0) than the control group (\textit{M} = 16.9). The interaction between time and condition was not significant (\textit{f}(1,23) = 1.03, \textit{p} = .32, \(\eta_p^2\) = .04), meaning that both groups showed a similar drop in the amount of feedback given in session 2.

\subsection{Does a GenAI assistance lead to higher-quality student feedback?}

To answer RQ2, the same dataset and statistical analyzes were used. Regarding the overall quality of the feedback, the treatment group outperformed the control group by an average of 1.5 points between the two sessions. This aggregate difference was statistically significant, as shown by a significant main effect of condition (\textit{f}(1,21) = 5.82, \textit{p} = .03, \(\eta_p^2\) = .22), indicating that students in the treatment group provided higher-quality feedback (\textit{M} = 5.6, out of a maximum score of 9) compared to those in the control group (\textit{M} = 4.1). These findings support our second hypothesis that a GenAI-driven gamified approach enhances feedback quality. There was no main effect of time (\textit{f}(1,21) = .06, \textit{p} = .81, \(\eta_p^2\) = .003), such that the overall quality of the feedback did not differ between session 1 (\textit{M} = 4.8) and session 2 (\textit{M} = 4.6). There was a significant interaction between time and condition (\textit{f}(1,21) = 4.61, \textit{p} = .04, \(\eta_p^2\) = .18).

When different dimensions of feedback quality are examined, the results become more nuanced. First, in relation to \textit{ clarity}, the feedback provided by the students in both groups was consistently high, without significant main effect of the condition ($f(1,21) = .07$, $p = .79$, $\eta_p^2 = .00$). Similarly, there were no significant differences between sessions, indicating that the students in both groups continued to excel in providing clear feedback. The high performance in this aspect suggests that the students maintained their ability to provide clear peer feedback regardless of the treatment. We hypothesize and explain this result in \Cref{discussion}.

In contrast, the \textit{relevance} of feedback (defined as comments that are relevant to content, presentation, or delivery) showed a significant main effect of the condition ($f(1,21) = 7.41$, $p = .01$, $\eta_p^2 = .26$). The treatment group provided more relevant feedback (\textit{M} = 1.8) compared to the control group (\textit{M} = 1.4). A similar pattern emerged for \textit{specificity}, where the treatment group also outperformed the control group. The mean specificity score for the treatment was 1.2, whereas the control scored only 0.4. This suggests that the GenAI-driven gamified treatment effectively motivated the students to provide feedback that was not only more relevant but also more detailed and precise. In particular, specificity proved to be the most challenging aspect for students to achieve, as evidenced by the low scores in the control group, which lacked external motivation to focus on this dimension.

An unexpected observation concerns the evolution of relevance and specificity scores over time in both sessions. Contrary to expectations that students in both groups would maintain or improve their performance, it was actually the control group that showed improvement from Session 1 to Session 2. Furthermore, the interaction between time (session) and condition was statistically significant for both aspects of quality: relevance ($f(1,21) = 9.14$, $p = .006$, $\eta_p^2 = .30$) and specificity ($f(1,21) = 5.64$, $p = .03$, $\eta_p^2 = .21$).

These interaction effects suggest that while the treatment initially improved feedback quality, the control group demonstrated a significant learning effect over time, slightly narrowing the gap between the groups in session 2.

In summary, while the GenAI-driven gamified approach significantly improved feedback quality in terms of relevance and specificity compared to the control group, there were notable temporal dynamics. The treatment had a solid initial effect, but the control group improved over time, emphasizing the potential for students to adapt and enhance their feedback skills without external motivation. This suggests that while gamified elements can effectively foster certain aspects of feedback quality, intrinsic factors, and natural progression also play important roles in student development over time. We also discuss this result in \Cref{discussion}.

\section{Discussion and Qualitative Findings}
\label{discussion}

The results in terms of the quantity and quality of peer feedback demonstrate the effectiveness of the GenAI-driven gamified approach. We provide qualitative information here to explain these findings. Although it is challenging to establish a direct relationship between specific gamification mechanics, GenAI assistance, and their integrated impact on the quantity or quality of feedback, the overall effect was significant. The integration of rivalry, accomplishment features, and GenAI assistance into the platform reassures us of the robustness of our approach. This successful integration encouraged the treatment group to provide multiple high-quality optional reviews, which improved the impact of peer feedback.

\subsection{Temporal Analysis of Engagement and Quality}
Upon data analysis (\Cref{tab:feedback_comparison}), a trend observed throughout the cohort was that students provided more feedback during the first session than during the second. This decline can be attributed to the timing of the second session, which coincided with the last week of classes, often referred to as ``crunch time,'' when students manage multiple assignments and deadlines in their courses. Despite these pressures, the treatment group consistently provided a greater amount of feedback than the control group throughout both sessions, supporting our hypothesis that gamification would lead to increased student engagement.

Another observation concerns the evolution of relevance and specificity scores in both sessions. Unlike expectations that both groups would maintain or improve their performance, the control group showed improvement over time. Furthermore, the interaction between time and condition was significant for both relevance ($f(1,21) = 9.14$, $p = .006$, $\eta_p^2 = .30$) and specificity ($f(1,21) = 5.64$, $p = .03$, $\eta_p^2 = .21$).

The timing of the second session (during ``crunch time'') could also explain this trend. Although the students in the treatment group were still motivated to provide more feedback, the quality of their reviews may have suffered slightly under pressure. However, the scores of the treatment group for all quality aspects remained equal to or higher than those of the control group, indicating that the intervention still had a positive effect.

\subsection{Student Engagement with GenAI Feedback Assistance: Survey Insights}

At the end of the semester, we asked the students to provide feedback on their experience using the \textit{Socratique} platform through a survey. The survey used a Likert scale, with ratings from 1 to 7, to capture students' perceptions of various aspects of the platform, including its usability, the impact of gamification elements, and the effectiveness of the AI feedback assistant, Mr. Pepper. This approach allowed us to collect quantitative information on how students interacted with \textit{Socratique}, their attitudes towards its characteristics, and the perceived benefits and challenges associated with using it for peer evaluation. \Cref{tab:survey} shows these quantitative results. 

\begin{table}[]
\resizebox{\columnwidth}{!}{%
\begin{tabular}{l|r|r|l}
\hline
\rowcolor[HTML]{C0C0C0} 
\multicolumn{1}{c|}{\cellcolor[HTML]{C0C0C0}\textbf{Survey Item}} & \multicolumn{1}{c|}{\cellcolor[HTML]{C0C0C0}\textbf{\begin{tabular}[c]{@{}c@{}}Control\\ Mean\\ (n = 14)\end{tabular}}} & \multicolumn{1}{c|}{\cellcolor[HTML]{C0C0C0}\textbf{\begin{tabular}[c]{@{}c@{}}Treatment\\ Mean\\ (n = 11)\end{tabular}}} & \multicolumn{1}{c}{\cellcolor[HTML]{C0C0C0}\textbf{\begin{tabular}[c]{@{}c@{}}Statistical \\ Result\end{tabular}}} \\ \hline
\begin{tabular}[c]{@{}l@{}}I provided honest and unbiased ratings to my\\ peers regardless of how that impacted their grades.\\ {[}1 = strongly disagree, 7 = strongly agree{]}\end{tabular} & 6.6 & 6.5 & \begin{tabular}[c]{@{}l@{}}\textit{t(23)} = 0.49, \\ \textit{p} = 0.63\end{tabular} \\ \hline
\begin{tabular}[c]{@{}l@{}}I felt motivated to give peer feedback on the platform.\\ {[}1 = strongly disagree, 7 = strongly agree{]}\end{tabular} & 5.9 & 5.6 & \begin{tabular}[c]{@{}l@{}}\textit{t(23)} = 0.64, \\ \textit{p} = 0.53\end{tabular} \\ \hline
\begin{tabular}[c]{@{}l@{}}How would you rate the quality of open-ended\\ feedback received from your peers on the platform?\\ {[}1 = very poor, 7 = very good{]}\end{tabular} & 6.0 & 5.9 & \begin{tabular}[c]{@{}l@{}}\textit{t(23)} = 0.28,\\ \textit{p} = 0.78\end{tabular} \\ \hline
\begin{tabular}[c]{@{}l@{}}How often did you consult the AI feedback assistant\\ for insights on the feedback you provided?\\ {[}1 = never, 7 = for every evaluation{]}\end{tabular} & 3.4 & 5.6 & \cellcolor[HTML]{C0C0C0}\begin{tabular}[c]{@{}l@{}}\textit{t(23)} = -3.34,\\ \textit{p} = 0.003\end{tabular} \\ \hline
\begin{tabular}[c]{@{}l@{}}Do you feel that the AI feedback assistant helped you\\ improve the quality of your open-ended feedback\\ (in terms of clarity/specificity/relevance)?\end{tabular} & 4.8 & 5.7 & \begin{tabular}[c]{@{}l@{}}\textit{t(23)} = -1.68,\\ \textit{p} = 0.11\end{tabular} \\ \hline
\begin{tabular}[c]{@{}l@{}}I found the peer-assessment experience on the\\ platform enjoyable.\\ {[}1 = strongly disagree, 7 = strongly agree{]}\end{tabular} & 5.6 & 5.6 & \begin{tabular}[c]{@{}l@{}}\textit{t(23)} = -0.17,\\ p = 0.86\end{tabular} \\ \hline
\begin{tabular}[c]{@{}l@{}}Using the platform helped me improve my skills\\ in giving feedback. \\ {[}1 = strongly disagree, 7 = strongly agree{]}\end{tabular} & 5.6 & 5.9 & \begin{tabular}[c]{@{}l@{}}\textit{t(23)} = -0.82,\\ \textit{p} = 0.42\end{tabular} \\ \hline
\end{tabular}%
}
\caption{Survey Results for Control and Treatment Groups}
\label{tab:survey}
\end{table}

The results of the end-of-semester survey (\Cref{tab:survey}) reflect the usage patterns of the platform by students and align with the peer feedback data. Although most of the questions did not reveal significant differences between the treatment and control groups, one question stood out: ``How often did you consult the AI feedback assistant to get insight into the feedback you provided?'' The intervention group reported significantly greater use of the GenAI assistant, Mr. Pepper, with an average score of 5.6 compared to 3.4 in the control group. Although points were awarded only for the initial interaction with Mr. Pepper, students in the gamified version of the platform continued to seek insight from the assistant, motivated by the desire to provide higher-quality feedback, which in turn was rewarded through badges and point multipliers when achieved consistently.

Another notable finding from the survey pertains to students' perceptions of the usefulness of the GenAI assistant in enhancing the quality of peer feedback and their skills in providing feedback. The treatment group, which consulted Mr. Pepper more frequently, rated his assistance as having a marginally higher impact on the quality of their open feedback, with an average score of 5.7, compared to 4.8 in the control group. A similar trend, but less pronounced, emerged concerning improvement in students' feedback sharing skills. This suggests that the increased engagement with the GenAI assistant positively affected students' perceptions of the feedback quality improvement process.

\subsection{Understanding the Population Characteristics}
While clarity might be the most straightforward aspect of improving feedback quality, the students in our study consistently excelled in this area, with no significant differences between the intervention and control groups. We believe that this is related to the study population.
As revealed by a recent literature review, peer feedback investigations are often carried out among undergraduate students \cite{indriasari2020review}, whereas our study concerned Master's students that exhibit specific characteristics. First, communication skills are strongly emphasized in our program, and all courses require students to give at least one presentation, where clarity is part of the assessment rubric. Second, due to the program's small size, students receive a considerable amount of personalized feedback from instructors, which helps foster good practices in communication. Thus, we believe that GenAI assistance in honing clarity may be more valuable in other populations, particularly among less experienced students (e.g., undergraduates) or in programs where public speaking opportunities and individualized feedback are more scarce.
\subsection{Feedback relevance, specificity, and the role of GenAI} 
When comparing different aspects of feedback quality, we observed that achieving relevance was more challenging than clarity, particularly for generic feedback, such as ``good job,'' ``nice,'' or ``I like it.'' The GenAI assistant helped address these shortcomings by encouraging students to focus on specific aspects of the presentation and provide richer and more constructive feedback. Another interpretation of this could be the timing of when feedback was entered into the system. Given that the students had a few days after observing the presentations, they may have forgotten what specifically made them think that it was a good presentation (if no notes were taken during the presentation). A future study could alter the time between observing presentations and providing peer feedback to see if the effects of the gamified system differ.

Specificity proved to be the most challenging quality to achieve, as it required students to closely follow the presentation and identify concrete strengths or areas for improvement, such as ``your speech volume was too low for students in the back'' or ``the results slide had too much text; using graphs would have been more effective.'' Providing specific feedback is not only labor intensive and, therefore, heavily dependent on student motivation, but also challenging for generative AI to improve upon. The challenge is twofold. First, in-depth evaluation of feedback relevance significantly increases processing time, as it involves parsing the entire presentation artifact rather than just a few lines of text. This can make real-time feedback impractical in a live setting. Second, the content presented is often minimal. Effective presenters tend to use concise key messages and visual aids, limiting the information available to GenAI to assess relevance effectively. Furthermore, aspects of the delivery, such as tone or body language, were completely inaccessible to the digital assistant, further constraining its potential to provide tutoring. These limitations apply to a smaller degree to other types of deliverables that can be subject to peer review - design documents, source code - and where Socratique could be employed.

\subsection{Motivational Rewards as a Key Driver}
Based on our experience and observations, establishing effective rewards was crucial to driving student motivation. In previous experiments, financial rewards, such as Amazon or restaurant vouchers, were used but did not seem particularly desirable to students. In this experiment, we opted for grade-based rewards, which had an increasing level of appeal. Although integrating the peer review process into the course grade has been previously employed \cite{clark2004peer, hundhausen2011online}, we posit that the success of gamification depends on offering substantial rewards. Specifically, we provided grade enhancements of up to 4\%, compared to the maximum 2\% reported in existing literature. This significant increase aimed to make the rewards more attractive, particularly towards the end of the semester, thereby encouraging sustained participation. 

To further maintain students' motivation to provide numerous high-quality peer reviews, we designed the point redemption system such that rewards could not be earned within a single presentation session (session 1). This strategic threshold was intended to promote continuous participation and consistent participation throughout the course. By requiring students to accumulate points over multiple sessions, we aimed to foster a habit of regular peer feedback, thereby enhancing both the quantity and quality of reviews.
 
\section{Limitations and Threats to Validity}
\label{sec:limitations}

Although the findings of this study provide valuable information on the effectiveness of a gamified approach driven by GenAI to improve the quantity and quality of peer feedback, we must acknowledge several limitations and potential threats to validity.

\subsection{Internal Validity}
One threat to internal validity is the timing of the intervention. The second session took place during the last week of classes, a period characterized by high academic pressure on the students. This timing may have affected student motivation and participation differently in different sessions, potentially influencing the quantity and quality of feedback given. As a result, it is difficult to fully disentangle the effects of the intervention from the external stressors experienced by students.

\subsection{External Validity}
The characteristics of the study participants limit the generalizability of our findings. The experiment was carried out with a nonrandom sample of students from a single institution and within a specific course, which may limit the applicability of the results to other educational settings or subjects. The effectiveness of the GenAI-driven gamified approach can vary significantly in different contexts, with different demographics of participants, or in larger-scale deployments.

In addition, the study relied on a voluntary participation model, which can introduce self-selection bias. Students who were more comfortable with technology or more motivated to provide peer feedback may have been more likely to engage with the gamified platform, potentially skewing the results in favor of the intervention. However, because the students were randomly assigned to the treatment or control group, we do not believe that this significantly impacted our study or the outcome of our gamification intervention.

The artifacts subjected to peer review in this study were limited to relatively short presentations. In real-world software engineering practices, peer reviews often involve more complex artifacts such as design documentation and source code, which require greater effort and time from reviewers. This difference in workload may limit the generalizability of our findings to such contexts. However, we believe that the gamification approach can scale effectively through strategic scheduling and adjustment of the number of mandatory reviews required.
\subsection{Construct Validity}
Construct validity could also be affected by the way feedback quality was operationalized and measured. The quality of the feedback was assessed using a rubric that focused on clarity, relevance, and specificity, each with a maximum score of 3. Although these dimensions are essential for feedback quality, other important aspects, such as depth of understanding or actionability, were not measured, which might limit the completeness of our evaluation.

Furthermore, operationalization of the ``quantity of feedback'' as the number of peer reviews provided does not capture potential differences in the depth or length of individual feedback comments. A student could offer several short, superficial comments that count towards the total quantity but may not reflect meaningful engagement with peer work.
\subsection{Conclusion Validity}
The study's sample size was relatively small, with only 34 participants in the treatment and control groups. This limited sample size reduces the statistical power of the analysis and increases the risk of Type II errors (failing to detect an effect that is present). Future studies should include a larger sample size to increase the robustness of the findings.

Finally, the study only included two sessions of peer feedback, which might need to be expanded to fully capture long-term learning effects or changes in student behavior. A longitudinal study with more sessions and follow-up assessments would provide a more comprehensive understanding of how a GenAI-driven gamified approach impacts feedback quality over time.

\section{Conclusion}
\label{sec:conclusion}

Peer feedback in software engineering (SE) courses serves as an early introduction to professional review practices, such as code reviews and design critiques. Our findings show that combining gamification with GenAI scaffolding can enhance student peer feedback outcomes, both in terms of feedback quality and self-perceived skill acquisition. In typical peer review processes, reviewers require motivation (to thoroughly inspect an artifact) and the ability to articulate actionable improvements; Socratique’s integrated approach addresses both fronts, offering a promising blueprint for cultivating feedback literacy in SE education. Our results demonstrate that the treatment group (students using gamified Socratique driven by GenAI) provided more peer feedback of higher quality than the control group, particularly in terms of relevance and specificity. In addition to improving engagement, the study also shows how GenAI can be used for \textit{ in-process evaluation} of peer feedback quality. By evaluating critiques in real time, Mr. Pepper offers formative guidance that students can apply immediately, mirroring iterative and incremental development practices found in industry.

Although our approach was successful for student presentations, the same principles can be adapted to other artifacts common in SE courses, ranging from design documentation and UML diagrams to source code, provided suitable prompts or fine-tuning align the LLM to domain-specific criteria. This paves the way for educators to scale peer assessment to more technical and detailed tasks without overburdening instructors. Ultimately, such a strategy can foster deeper engagement, reinforce industry-aligned review behaviors, and strengthen overall student readiness for collaborative software engineering.
We also acknowledge several limitations, such as the potential influence of the timing of the intervention and the constraints of a small sample size, all of which may limit the generalizability of our findings. This candid recognition of the study's limitations underscores our commitment to rigorous research. Future research should focus on addressing these limitations by involving larger, more diverse samples and employing longitudinal designs to evaluate long-term impacts.

In conclusion, our findings, which represent the first study to include GenAI in a gamified peer feedback intervention, provide significant evidence that gamification, when integrated with GenAI, can significantly enhance the effectiveness of peer feedback in educational environments. This approach has the potential to improve student engagement and learning outcomes by fostering a more interactive and motivating feedback process. However, more research is required to replicate these results in diverse educational contexts, various types of SE artifacts, and to explore the long-term impacts of such interventions on student learning and collaboration. The findings of our study, therefore, shed light on a promising avenue to improve educational practices.

\bibliographystyle{ACM-Reference-Format}
\bibliography{references.bib}

\end{document}